\documentclass[12pt]{article}
\usepackage{amsmath,amssymb,exscale}  
\input epsf
\def\gappeq{\mathrel{\rlap {\raise.5ex\hbox{$>$}}
{\lower.5ex\hbox{$\sim$}}}}
\def\permil{$\%\raise.20ex\hbox{$_0$}$}
\def\lappeq{\mathrel{\rlap{\raise.5ex\hbox{$<$}}
{\lower.5ex\hbox{$\sim$}}}}
\newcommand {\nn} {\nonumber}
\newcommand {\half} {\frac{1}{2}}

\begin{document}
\topmargin -0.8cm
\oddsidemargin -0.8cm
\evensidemargin -0.8cm
\pagestyle{empty}
\begin{flushright}
SNUTP-00-022\\
hep-th/0008190
\end{flushright}
\begin{center}
{\large\bf Consistency of Higher Derivative Gravity in the Brane Background}\\
\vspace{0.6cm}
{\large Ishwaree P. Neupane\footnote{Email: ishwaree@phya.snu.ac.kr}}\\
\vspace{0.4cm}
{\it{Department of Physics, Seoul National University, 151-742, Seoul, 
Korea}}\\
\end{center}
\vspace{1cm}
\begin{abstract}
We consider the theory of higher derivative gravity with non-factorizable 
Randall-Sundrum type space-time and obtain the metric solutions 
which characterize the $p$-brane world-volume as a curved or planar defect 
embedded in the higher dimensions. We consider the string inspired effective 
action of the dilatonic Gauss-Bonnet type in the brane background and show 
its consistency with the RS brane-world scenario and the conformal 
weights of dilaton couplings in the string theory with appropriate 
choice of Regge slope ($\alpha'$) or Gauss-Bonnet coupling ($\alpha$) 
or both. We also discuss time dependent dilaton solutions 
for a version of string-inspired fourth-derivative gravity model. 
\end{abstract}
\begin{flushleft}
SNUTP-00-022\\
August 2000
\end{flushleft}
\vfill
\eject
\pagestyle{empty}
\setcounter{page}{1}
\setcounter{footnote}{0}
\pagestyle{plain}
\section{Introduction}
Initially, the higher dimensional gravity theories \cite{BSD,CWE} 
were realized with the assumptions that the realistic Kaluza-Klein theories
\cite{APP} require compact extra spatial dimensions, other than 
$1+3$ space-time world, of the order of Planck lengths. However, this idea 
has got a new interest and direction after the intriguing ideas of Randall
and Sundrun (RS) that the physical $3+1$ dimensional space-time world can 
be embedded as a hypersurface in a higher dimensional anti-deSitter ($AdS$) 
bulk space-time \cite{RS1}, leaving the extra non-compact spatial dimensions 
\cite{VAR} of the size of $mm$ or even infinite \cite{RS1,NAH}. What is 
observed crucial about this brane scenario is that at shorter distance than 
the $AdS$ curvature length, the extra dimensions reveal to change the 
Einstein gravity significantly and above the $AdS$ radius it is effectively 
approximated by the Einstein's gravity in $3+1$ space-time dimensions~ 
\cite{RS1,CCS}, pointing the fact that the correction is much suppressed 
as the number of transverse extra dimensions grow \cite{TGM}. This has left 
room for the gravity (and possibly the spin zero counterpart of graviton, 
i.e. dilaton) to flow freely in the bulk, whereas the standard matter 
particles are effectively confined in the brane. Such localization of matter 
on the brane \cite{RS1,TGM,ODA} has changed our conventional wisdom on 
KK theories \cite{APP}, i.e. the size of extra dimensions does not have to 
be $~1/M_{pl}\sim 10^{-33} cm$, but could be much bigger.

In most of recent work [8-14] to the solution of Einstein equations in 
$n\geq 1$ extra dimensions, the $3+1$ dimensions of our world is 
identified with the internal space of topological 
defects residing in a higher dimensional space-time. Specifically, 
the $3$-brane world volume is pictured as a domain wall propagating in the 
five-dimensional bulk space-time \cite{RS1,ODA}, as a local string defect 
residing in the six-dimensional space-time \cite{TGM,ODA,ACE,AGC}, as a 
global monopole defect with a conical deficit angle in seven-dimensional 
space-time \cite{IAV} and its continuation with the co-dimensions four 
characterizes the instanton solution which describes the quantum nucleation 
of a five-dimensional brane-world \cite{JGM,IAV}. Most of these and similar 
work have been judged only from the solutions to Einstein gravity in the 
extra dimensions. Also in \cite{LCK} an idea about the creation of a 
spherically symmetric brane-world for a $1$-brane and a $3$-brane in the 
context of Wheeler-De Witt equation and WKB approximation is furnished. 
Thus, in this paper, we shall find metric solutions that also include 
the contribution from higher-curvature terms in the field equations and 
describe the $p$-brane object as a curved or planar defect residing in the 
higher dimensions.

In order to check the compatibility of the various string inspired low 
energy effective actions under the brane scenarios and to realize the very 
natural and more fundamental theory of gravity one cannot avoid the 
higher-curvature terms in the effective action. This is so from the 
different perspectives such as the renormalizibility and the asymptotic 
freedom of the theory in 4 dimensions \cite{KKS}, and for the stabilization 
of the scalar potential \cite{CWE} in the bulk. Also the action 
containing second powers of the curvature tensors naturally arise in the 
string effective action \cite{GSW}. This has further attraction with the 
hope that some versions of the higher derivative gravity represent the 
supergravity duals when $R$ and $\Lambda$ are the leading order and higher 
derivatives are of next to leading order in large $N$ expansion 
(see \cite{NOJ} and references therein). The effective action containing 
just $R$ cannot stabilize the scalar potential defined in the bulk, 
otherwise, the bulk gauge sector appears in conflict with the large extra 
dimensions \cite{CWE,NAH}.

An investigation of the higher derivative gravity by adding a dilatonic 
potential to the effective action is made in ref.\cite{ILZ} where a naked 
singularity in $AdS_5$ bulk persists for the finite $3+1$ dimensional Planck 
mass and dilaton potential acts as the bulk cosmological constant. Also 
various static and inflationary solutions of the RS model in the Gauss-Bonnet 
combination have been studied in \cite{KKL}. But, what is missing in these 
models is either dilaton or the proper coupling of dilaton to the 
higher-curvature terms or both. So we consider the string inspired effective 
action \cite{AAT,ESF} in dilatonic Gauss-Bonnet combination taking the 
proper care of conformal weights of the dilaton couplings. A work along 
this line is recently reported in \cite{NIC}. 

The paper is organized as follows. In the next Section we give the 
solutions to higher derivative gravity in the background of Ricci constant 
curvature space-time. In Section 3 we derive the Einstein field equations 
for a non-factorizable general metric ansatz and will be subsequently 
utilized in Section 4 to obtain the metric solutions corresponding to higher 
dimensional defects. In Section 5 we also include the contribution of the 
higher curvature terms to the field equations and find the metric solutions 
that characterize planar wall and planar string defects. 
In Section 6 we consider string inspired effective actions in the dilatonic 
Gauss-Bonnet combination and study compatibility of the theory with 
RS type space-time and the conformal weights of dilaton couplings derived  
from heterotic type I string theory in five space-time dimensions. 
Our discussions and outlook of the problems are summarized in Section 7.
\section{Higher Derivative Gravity}
The most general effective action involving invariants with mass dimension 
$\geq 4$ and vanishing torsion and dilaton field for the higher derivative 
gravity is given by
\begin{equation}
S=\int d^{d+1}x\sqrt{-g}\Big\{\frac{R}{2\kappa^2}-\Lambda_b
+\lambda'\big(\alpha R^2+\beta R_{AB}R^{AB}+\gamma R_{ABCD}R^{ABCD}\big)
\Big\}+S_B+S_m
\end{equation}
where $\Lambda_b$ is the bulk cosmological constant, $S_B$ is the boundary 
action and $S_m$ is the matter action which may include the matter trapped
 in the $p$-brane and the $D(=d+1)$ dimensional mass term is defined by 
$\kappa^2 = 8 \pi G_{d+1} = 8\pi M^{1-d}$. Here we set the string coupling 
$\lambda'=\alpha'/{8g_s^2}=1$ for simplicity and but the effective action 
with the proper couplings of dilaton field to the higher-curvature 
terms and $\lambda'>0$ will be presented in Section 6.

The classical field equations derived by varying the above action w.r.to 
$g^{AB}$ can then be expressed in the form 
\begin{equation}
G_{AB}+\kappa^2 X_{AB}=-\kappa^2(\Lambda_b g_{AB}-T_{AB})
\end{equation}
where the energy-momentum tensor is defined in the well known form
\begin{equation}
T_{AB}=-\frac{2}{\sqrt{-g}}\frac{\delta S_m}{\delta g^{AB}}
=-\frac{2}{\sqrt{-g}}\frac{\delta}{\delta g^{AB}}\int d^{d+1}x 
\sqrt{-g}~{\cal L}_m,
\end{equation}
$G_{AB}=R_{AB}-g_{AB}R/2$ and $X_{AB}$ is given by
\begin{eqnarray}
X_{AB} &=&-\frac{1}{2}g_{AB} (\alpha R^2+\beta R_{CD}R^{CD}+
\gamma R_{CDEF}R^{CDEF})\nn\\
& &+2\big[ \alpha R R_{AB}+\beta R_{ACBD}R^{CD} +
 \gamma (R_{ACDE}R_B\,^{CDE}-2R_A\,^C R_{BC}+2R_{ACBD}R^{CD})\big]\nn\\
& &-(2\alpha+\beta+2\gamma)(\nabla_A\nabla_B R - g_{AB}\nabla^2R)+
(\beta+4\gamma)\nabla^2G_{AB}
\end{eqnarray}
where $A,B,\cdots$ denote the $D=d+1$ dimensional space-time indices and 
the signature we follow is $(-,+,+,+,\cdots)$.  
The last line of the above expression vanishes for two cases i.e. 
(i) for  the Gauss-Bonnet relation (i.e. $4\alpha=-\beta=4\gamma$) and 
(ii) for the background metric of Ricci constant curvature space-time 
(in which case, since $R_{AB}\propto g_{AB}$ and the curvatures are
 covariantly constant). To simplify the calculation to a greater extent, 
we keep the freedom in the choice of relations among $\alpha,\beta$ 
and $\gamma$ for the later purpose and  mainly consider the second case in 
what follows in this section.
We briefly review the general features of the action (1) at first. The 
general curvature squared terms in the $d$-spatial dimensions can be 
written as \cite{ILS},
\begin{eqnarray}
\alpha R^2+\beta R_{AB}R^{AB}+\gamma R_{ABCD}R^{ABCD}\nn
\end{eqnarray}
\begin{equation} 
=-\bigg[\frac{(d-1)\beta+4\gamma}{4(d-2)}\bigg]{\cal R}_{GB}^2
+\bigg(\frac{d-1}{d-2}\bigg)(\beta /4+\gamma)C^2
+\bigg[\frac{4d\alpha+(d+1)\beta+4\gamma}{4d}\bigg]R^2  
\end{equation}
where the Gauss-Bonnet term $({\cal R}_{GB}^2)$ and the square of the 
Weyl tensor $(C^2)$ are given by
\begin{eqnarray}
{\cal R}_{GB}^2 &=& R^2-4R_{AB}R^{AB}+R_{ABCD}R^{ABCD},\nn\\
C^2 &=&\frac{2}{d(d-1)}R^2-\frac{4}{d-1}R_{AB}^2+R_{ABCD}R^{ABCD}
\end{eqnarray}
Ofcourse, for $d=3$, ${\cal R}_{GB}^2$ is simply the integrand of the Gauss-Bonnet term.
Since the Weyl tensor vanishes for a conformally flat $AdS$ metric, with the 
requirement that $16\alpha+5\beta+4\gamma=0$ for $d=4$, the curvature 
squared terms naturally arise in the Gauss-Bonnet combination. 
In the scheme where $R_{ABCD}^2$ can be modified to $C_{ABCD}^2$ the 
action (1) may be deduced from the heterotic string via heterotic type
 I duality \cite{AAT,ESF}. Furthermore, the theory with the Weyl term squared 
corresponds to the setting $\alpha=-\beta/{2d}=2\gamma/{d(d-1)}$. 
For $d=4$ this means $\alpha=\gamma/6,\beta=
-4\gamma/3$ and; evidently, the higher derivative part of the 
above action is reduced to
\begin{equation}
S=\int d^{d+1}x \sqrt{-g}\Big(\frac{R}{2\kappa^2} -\Lambda_b+\gamma
 C_{ABCD} C^{ABCD}\Big)+S_m+S_B
\end{equation}
With $T_{AB}=0$, the theory defined by the above action (7) can have 
only the flat 
space-time as the vacuum solution. Forthermore, the 
 Einstein gravity with the cosmological constant term is characterized 
by the setting $\alpha=\beta=\gamma=0$, i.e.,
\begin{equation}
S=\int d^{d+1}x\sqrt{-g}\Big(\frac{R}{2\kappa^2}-\Lambda_b\Big)
+\int d^dx\sqrt{-g}{\cal L}_m+S_B
\end{equation}
This gives (with $S_B=0$ and $T_{AB}=0$)~$R_{AB}-g_{AB}R/2=
-\kappa^2 \Lambda_b g_{AB}$ and the corresponding $d+1$ dimensional Ricci 
curvature and Ricci scalar are then given by
\begin{equation}
R_{AB}= -\frac{d}{l^2}g_{AB},\,\, R=-\frac{d(d+1)}{l^2}
\end{equation}
where the square of the curvature radius is defined by 
$l^2=-d(d-1)/2\kappa^2 \Lambda_b$. For $d=3$, this is the square 
of the curvature radius in the Minkowski space-time. In our notations 
$AdS(dS)$ solution corresponds to $\Lambda_b<0, l^2>0 (\Lambda_b>0, l^2<0)$.
We assume that, for $d+1=5$, the equations of motion derived from (1) can 
have a solution describing the $AdS_5$ spacetime. This is indeed the 
case and whose metric solution is given by \cite{NOJ},
\begin{equation}
ds^2=\frac{1}{r}\eta_{\mu\nu}dx^{\mu}dx^{\nu}+\frac{l^2}{4r^2}dr^2
\end{equation}
This is actually a conformally flat $AdS$ metric solution. 
For $AdS_5$, the scalar, 
Ricci and Riemann curvatures (derived from the metric (10) or the 
eqn(9)) are given by $(l^2>0)$,
\begin{equation}
R=-\frac{20}{l^2},~ R_{AB}=-\frac{4}{l^2}g_{AB},~ R_{ABCD}=-\frac{1}{l^2}
(g_{AC}g_{BD}-g_{AD}g_{BC})
\end{equation}
This implies that the vacuum solution of Einstein gravity is equivalent
to the higher derivative gravity theory with the Ricci constant curvature bulk
 $AdS$. One then easily evaluates 
\begin{eqnarray}
R_{ACDE} R_{B}\,^{CDE}=-\frac{2d}{l^4} g_{AB}, 
 R_{ABCD} R^{ABCD}=-\frac{2d(d+1)}{l^4}
\end{eqnarray}
This further shows that the space-time of constant Riemann curvature 
characterizes the $AdS_{d+1}$ with $\Lambda_b<0$ and $l^2>0$.
The relations(11-12) are equally applied to the $(d+1)$ dimensional
conformally flat metric solution given by
\begin{equation}
ds^2=\frac{1}{r}\eta_{\mu\nu}dx^{\mu}dx^{\nu}+\frac{l^2}{4r^2}dr^2
+\frac{1}{r} d^{n-1}X^2
\end{equation}
where $d^{n-1}X^2$ denotes the $(n-1)$ dimensional flat space metric. 
With the above subtitutions into the equation (2) with $T_{AB}=0$, 
the field equations reduce to the following form:
\begin{equation}
\frac{d(d-1)}{2\kappa^2 l^2}-\frac{d(d-3)}{l^4} \big[ d(d+1)\alpha+
d\beta+2\gamma \big]+\Lambda_b=0
\end{equation}
There exists two classes of general solutions of the eqn(14) i.e. 
(i) for $l^2\to\infty$ : the solution is a flat Minkowski space-time with 
$\Lambda_b\to 0$ and
(ii) for the vanishing bulk cosmological constant, $(\Lambda_b=0)$, one has
\begin{equation}
l^2=\frac{2(d-3)}{d-1}\big[d(d+1)\alpha+d\beta+2\gamma\big]\kappa^2
\end{equation}
therefore, for $d>3$, there exist $AdS(dS)$ solution if the quantity
in the parenthesis is positive (negative). However for other values of 
$d$, viz $d\leq 3$, the case is less interesting. Specifically, the curvature 
diverges for $d=3$ and $d=1$ characterizes a flat space solution. For $d=2$, 
the space-time would be $dS (AdS)$ when the quantity in the parenthesis is 
positive (negative). The eqn(14) is solved for $l^2$ if,
\begin{equation}
\Big(\frac{d(d-1)}{2\kappa^2}\Big)^2+
4d(d-3)\left[d(d+1)\alpha+d\beta+2\gamma\right]\Lambda_b>0
\end{equation}
A connection of this result to the Supergravity dual of 
${\cal N}=2 Sp(N)$ and superconformal field theory is discussed in 
\cite{NOJ}. With the Gauss-Bonnet relation, 
i.e. $\alpha=-\beta/4=\gamma$, eqn(16) implies 
\begin{equation}
\alpha>-\frac{d(d-1)}{16\Lambda_b(d-2)(d-3)} \frac{1}{\kappa^4}
\end{equation}
This implies that for $d>3$ and $\alpha >0$, the bulk space-time is always 
$AdS$. And for the free parameters $\alpha, \beta, \gamma$, the space-time 
can be guaranted as $AdS_{d+1}$ if the following inequality holds,
\begin{equation}
4\Lambda_b d(d-3)\big[d(d+1)\alpha + d \beta + 2 \gamma \big] < 0
\end{equation}
\section{Einstein Field Equations}   
   In this section we  consider a general non-factorizable metric ansatz 
which might respect the $4d$ Poincare invariance. More specifically, we 
consider the presence of a brane and gravity. Due to the gravitational 
field on the brane, the brane can be curved and this would eventually 
rise to give a non-zero cosmological constant in usual $3+1$ space-time 
world. A $(d+1)$ dimensional metric ansatz satisfying this 
assumption is
\begin{eqnarray}
ds^2 &=& g_{AB} dx^A dx^B 
= g_{\mu\nu} dx^\mu dx^\nu +\gamma_{ij} dx^i dx^j\nn\\
&=& e^{-2M(r)}\hat{g}_{\mu\nu}dx^\mu dx^\nu + e^{-2N(r)} (dr^2 
+ d \Omega_{n-1}^2),
\end{eqnarray} 
where $\mu, \nu,\cdots$ denote the space-time indices of 
the $p$-brane, and $i,j,\cdots$ denote $n$-dimensional extra spatial 
indices. We also parameterize the $n-1$ dimensional sphere 
recursively as
\begin{equation}
d\Omega_{n-1}^2 = d\theta_{n-1}^2 + \sin^2 \theta_{n-1} d\Omega_{n-2}^2  
\end{equation}
So the equality $d=p+n$ holds and in this paper we concentrate on 
the case of $p\geq 3$. For simplicity one can reparametrize the ansatz
with the transformation $dy=e^{-N(r)} dr$ so that 
\begin{equation}
ds^2= e^{-2U(y)}\hat{g}_{\mu\nu}dx^\mu dx^\nu + dy^2 
+ e^{-2V(y)} d \Omega_{n-1}^2
\end{equation}
where $y$ is the radial bulk coordinate. Ofcourse, for $d+1=5$ and 
$\hat{g}_{\mu\nu}=\eta_{\mu\nu}$, $U(y)=(1+ k|y|)$ would explain the 
RS type solution with $AdS_5$ bulk geometry , where $k$ has the inverse 
dimension of $AdS_5$ curvature radius.
Variation of the action (8) with respect to the $(d+1)$ dimensional
metric tensor $g_{AB}$, with $S_B=0$, leads to the Einstein's equations,
\begin{equation}
R_{AB}-\frac{1}{2}g_{AB}R=-\kappa^2\left(\Lambda_b g_{AB}-T_{AB}\right)
\end{equation}
Here, we are interested to consider a spherically symmetric ansatz 
for the energy-momentum tensor in the form
\begin{eqnarray}
T^\mu_\nu &=& \delta^\mu_\nu\rho_o(y),~ T_y^y =\rho_y(y),~ \nn\\
T^{\theta_1}_{\theta_1} &=& T^{\theta_2}_{\theta_2} = \cdots =
T^{\theta_{n-1}}_{\theta_{n-1}} = \rho_\theta(y),
\end{eqnarray}
where $\rho_i (i=0,y,\theta)$ are the brane sources defined in a general 
$(d+1)$ dimensional space-time. With the metric ansatz (21), we evaluate the
followings non trivial components of the Einstein tensor, 
\begin{eqnarray}
G_t^t&=&-G_{x_1}^{x_1}=-G_{x_2}^{x_2}=-G_{x_3}^{x_3}
=-\Lambda_p e^{2U}+p\Big[(n-1)U'V'-U''+\frac{(p+1)}{2}(U')^2\Big]\nn\\
 && ~~~~~~~~~~~~~~~~~~~~~~~~~~~~~  
-\frac{n-1}{2}\Big[2V''-n(V')^2+(n-2)e^{2V}\Big]
\end{eqnarray}
\begin{equation}
G_y^y=-\frac{p+1}{p-1}\Lambda_p e^{2U}
+\frac{p+1}{2}\Big[2(n-1) U' V' + p (U')^2\Big] 
+\frac{(n-1)(n-2)}{2}\Big[(V')^2-e^{2V}\Big] 
\end{equation}
\begin{eqnarray}
G_{\theta_i}^{\theta^i}&=&-\frac{p+1}{p-1}\Lambda_p e^{2U}
+\frac{p+1}{2}\Big[(p+2)(U')^2-2U''+ 2(n-2)U'V'\Big]
-\frac{(n-2)(n-3)}{2}e^{2V}\nn\\
&& ~~~~~~~~~~~~~~~~~~~~~~~~~~~~~~~  -\frac{n-2}{2}\Big[2V''-(n-1)(V')^2\Big]
\end{eqnarray}
In general, the Ricci scalar corresponding to the metric ansatz (21) is
\begin{eqnarray}
R=\frac{2(p+1)}{p-1} \Lambda_p e^{2U}
+(p+1)\big(2U''-(p+2)(U')^2-2(n-1)U'V'\big)\nn\\
~~~~~~~+(n-1)\big\{2V''-n(V')^2+(n-2)e^{2V}\big\}
\end{eqnarray}
where the $'$ denotes the differentiation $d/dy$. In deriving the above
equations we have utilized the $(p+1)$ dimensional Ricci scalar,
$\hat R=2\Lambda_p(p+1)/(p-1)$ obtained from the $(p+1)$ dimensional 
Einstein equation   
\begin{eqnarray}
\hat{R}_{\mu\nu} - \frac{1}{2} \hat{g}_{\mu\nu} \hat{R} 
= - \Lambda_p \hat{g}_{\mu\nu}
\end{eqnarray}
Here we define the cosmological constant on the $p$ brane by the relation 
$\Lambda_p=8\pi\Lambda_{phy} M_{pl}^{p-1}$, and $\Lambda_{phy}$ is the 
physical cosmological constant defined in usual $(3+1)$ space-time. By 
eliminating two of the above equations or equivalently by using the 
energy momentum conservation law, $\nabla^A T_{AB}=0$, we get
\begin{eqnarray}
(p+1)(\rho_y-\rho_0)U'+(n-1)(\rho_y-\rho_{\theta})V'=\rho_y'
\end{eqnarray}
which is just the $(d+1)$ dimensional Bianchi identity derived from the 
Einstein field equations, i.e., after substituting (24-26) and (23) into (22).
\section{Global Defect Solutions}
   For the completeness we briefly discuss on the Eintein gravity 
(i.e. $\alpha=\beta=\gamma=0$) in higher dimensions, $n\geq 1$, and identify 
the $p$ brane world volume as the internal space of topological defect 
residing in the higher dimensional space-time.  
\subsection{Domain Wall Solution}
   For the codimension one, $n=1$, the Einstein field equations (24-25), with 
the ansatz (23), reduce to the following form
\begin{eqnarray}
2pU''-p(p+1)(U')^2&=&2\kappa^2[\Lambda_b-\rho_0(y)]-2\Lambda_p e^{2U}\\
-p(p+1)(U')^2&=& 2\kappa^2[\Lambda_b-\rho_y(y)]-
\frac{2(p+1)}{p-1}\Lambda_p e^{2U}
\end{eqnarray}
and the Bianchi identity(29) turns to the simpler form
\begin{equation}
\rho_y'=(p+1)(\rho_y-\rho_0)U'
\end{equation}
Now we consider the geometry having a warp factor, that is,
$U(y)=u|y|$, where $u$ is some constant. This ensures the localization of 
gravity to the $p$-brane.  With this ansatz, one easily solves 
eqns (30-31) to get the following general metric solution \cite{RS1, ODA}
\begin{equation}
ds^2=e^{-2u|y|}\hat g_{\mu\nu}dx^{\mu}dx^{\nu}+dy^2
\end{equation}
where
\begin{eqnarray}
\hat R_{(p+1)}&=&(p+1)\kappa^2 (\rho_0-\rho_y) e^{-2u|y|}\\
u^2&=&\frac{\kappa^2}{p(p+1)}\big[(p+1)\rho_0+(p-1)\rho_y-2\Lambda_b\big]
\end{eqnarray}
For the physically interesting case of $3$-brane, one has the
following general solutions
\begin{eqnarray}
\hat R_{(4)}&=&4\Lambda_p=4\kappa^2(\rho_0-\rho_y) e^{-2u|y|}\\
u^2&=&\frac{\kappa^2}{6}(2\rho_0-\rho_y-\Lambda_b)
\end{eqnarray}
There exists two special cases of these general solutions: i.e.
(i) for the trivial sources in the extra dimensions (i.e. $\rho_0=\rho_y=0$),
one has
\begin{equation}
\hat R_{(4)}=\frac{32\pi\Lambda_{phy}}{M_{pl}^2}=0,\,
u=\sqrt{-\frac{4\pi \Lambda_b}{3 M_{(5)}^3}}
\end{equation}
and hence the solution with $\Lambda_b<0$ ensures the bulk geometry as 
$AdS_5$ and the brane geometry as Ricci flat with $\Lambda_{phy}=0$, 
and (ii) the constant source terms in the extra dimensions 
(i.e. $\rho_0=\rho_y=const$) correspond to the spontaneous symmetry 
breaking in the extra dimensions. For the second case one has
\begin{equation}
\hat R_{(4)}=0, \, u=\sqrt{\frac{4\pi(\rho_0-\Lambda_b)}{3 M_{(5)}^3}}
\end{equation}
Clearly, $\rho_0>\Lambda_b$ guarantees the positivity of $u$ and 
exponentially decreasing warp factor needed to explain the hierarchy 
between electroweak scale and Planck scale mass \cite{NAH,RS1} and the fine 
tuning condition $\rho_0=\Lambda_b$ characterizes a flat 5d Minkowski 
space-time solution.
\subsection{String-like Solution}
   For the case of $n=2$ the Einstein field equations reduce to the form 
\begin{eqnarray}
V''- (V')^2-p u V'- \frac{p(p+1)}{2} u^2 
&=&\kappa^2 (\Lambda_b-\rho_0)-\Lambda_p e^{2u|y|}\\
-(p+1) u V'-\frac{p(p+1)}{2} u^2 
&=& \kappa^2(\Lambda_b-\rho_y)-\frac{(p+1)}{(p-1)}\Lambda_p e^{2u|y|}\\
-\frac{(p+1)(p+2)}{2} u^2
&=& \kappa^2(\Lambda_b-\rho_{\theta})-\frac{(p+1)}{(p-1)}\Lambda_p e^{2u|y|}
\end{eqnarray}
and the Bianchi identity (29) yields
\begin{equation}
\rho_y'=u(p+1)(\rho_y-\rho_0)+(\rho_y-\rho_{\theta})V'
\end{equation}
Solving the above field equations, the general metric solution can be
expressed in the form
\begin{equation}
ds^2= e^{-2u|y|}\hat{g}_{\mu\nu}dx^\mu dx^\nu + dy^2 
+ e^{-2V(y)} d\theta^2
\end{equation}
where
\begin{eqnarray}
u^2&=&\frac{1}{2(p+1)(p+2)}\big\{4\kappa^2(\rho_{\theta}-\Lambda_b)
+\hat R_{(4)} e^{2u|y|}\big\}\\
V(y)&=& u y +\frac{\kappa^2}{u(p+1)}\int dy (\rho_y-\rho_{\theta})\\
\hat R_{(4)}&=&\frac{\kappa^2}{u}\big(\alpha-\rho_{\theta}\big)e^{-2u|y|}
\end{eqnarray}
where $\alpha$ is an integration constant. Notice that the last equation 
above demands a definite form for $\rho_{\theta}$. One crucial observation
is that the inequality $\alpha > \rho_{\theta}$ keeps $\hat{R}_{(4)}$ 
positive and $(\rho_{\theta}-\Lambda_b)\geq 0$ ensures the positivity of 
$u^2$. Obviously, unlike the case of $3$-brane in domain wall solution, 
where only the case $\Lambda_b<0$ ensures $u^2>0$, for the string like 
solution one can either have $+ve$ or $-ve$ bulk cosmological constant 
to guarantee $u^2>0$ if $(\rho_{\theta}-\Lambda_b)>0$ still holds. It is 
also reported in \cite{ODA} that for the trivial sources in the extra 
dimensions, (i.e. $\rho_i=0$), only $\Lambda_b<0$ can ensure $u^2>0$. 
But this is not always the case as long as we do not set the integration 
constant $\alpha=0$. As a specific solution as $-\rho_y=\rho_{\theta}=
\alpha=constant$, spontaneous symmetry breaking condition in the extra 
dimensions \cite{RGR,ODA}, one obtains
\begin{equation}
ds^2= e^{-2u|y|}\hat{g}_{\mu\nu}dx^\mu dx^\nu + dy^2 
+ r_0^2 e^{-2u_0|y|} d\theta^2
\end{equation}
\begin{eqnarray}
u^2&=&\frac{2\kappa^2}{(p+1)(p+2)}(\rho_{\theta}-\Lambda_b)>0\nn\\
u_0&=&u-\frac{2\kappa^2 \rho_{\theta}}{u(p+1)},~~ \hat{R}_{(4)}=0    
\end{eqnarray}  
\subsection{Global Monopole-like Defect}
The above procedure may not directly apply to $n\geq 3$, because for $n=1$ 
and $n=2$ the extra space is either flat or conformally flat, but for $n\geq 
3$ the extra space is curved and hence the defect solution essentially 
introduces a solid angle deficit in the extra dimensions. The case is true 
even if one includes the higher-curvature terms 
into the field equations and assume $\hat{R}_{(4)}=0$. As gestured in 
\cite{DVA} the location of such defect in the brane might act as window 
in the higher dimensions. In the conventional Kaluza-Klein theory, our 
universe in higher dimensions has the topology of $M_4\otimes K$, where $K$ 
is the compact manifold and the isometries of $K$ can be seen as gauge 
symmetries of the effective four dimensional theory \cite{YMC} and
the $5$ dimensional space-time manifold is factorized as $M_4\otimes S^1$.
However, in the brane set up with a non-factorizable geometry, since the role 
of the gauge fields is played by the extra components of the graviton 
\cite{NAH}, the gauge fields may get mass when at least part of 
the $K$ isometries spontaneously break down in the extra dimensions by 
the brane. This effect can be observed as the Higgs phenomena 
in the usual $3+1$ dimensional world \cite{DVA}. This may suggest that a 
hedge hog type scalar field may explain the monopole defect in 
the extra dimensions. This is indeed the case as we see in ref.\cite{IAV} 
and also qualitatively discussed in \cite{DVA}.

One can consider the case of $n\geq3$ by defining the brane 
sources in terms of the radial hedge-hodge type scalar fields as in 
ref.\cite{IAV} and obtain metric solutions that characterize the monopole-like 
defects with a solid angle deficit in the extra dimensions. 
With this in mind, the source terms can be described by a multiplet of 
$n$ scalar fields $\Phi^a=\upsilon y^a/y$, 
where $\upsilon$ is the VEV at the minimum of the $n$-sphere, $y$ is the 
extra radial coordinate. Since $\rho_0(y)=\rho_y(y)=-(n-1)\upsilon^2/{2y^2}$ 
and $\rho_{\theta}(y)=-(n-3)\upsilon^2/{2y^2}$, the identity (29) 
implies that $V'(y)= -1/y$. This brings the non trivial term $e^{2V}$ to the 
simple form and hence one can solve the Einstein equations explicitly by 
making an appropriate ansatz for $U(y)$. As this paper mainly focuses on 
the dynamics with higher order curvature terms, an investigation of such 
defect solutions deserves as a topic for separate publication.
\nopagebreak 
\section{Higher Derivative Field Equations}
   In this section we shall confine ourselves to the case where the geometry
is Gauss-Bonnet type (i.e. $4\alpha=-\beta=4\gamma$) with the vanishing $3+1$ 
dimensional cosmological constant. We choose our metric ansatz in the form 
\begin{equation}
ds^2=e^{-2U(y)}\eta_{\mu\nu}dx^{\mu}dx^{\nu}+dy^2+e^{-2V(y)}d\Omega_{n-1}^2
\end{equation}
Using the Gauss-Bonnet relation, one can rewrite $X_{AB}$ from eqn(2) in the 
following form
\begin{equation}
X_{AB}=2\alpha(Y_{AB}-\frac{1}{4}g_{AB}{\cal R}_{GB}^2)= 2\alpha L_{AB}
\end{equation}
where $Y_{AB}$ is expressed by
\begin{equation}
Y_{AB}=R R_{AB}-2R_{ACBD}R^{CD}+R_{ACDE}R_B\,^{CDE}-2R_A\,^C R_{BC}
\end{equation}
In $3+1$ dimensions $L_{AB}$ is the covariantly conserved Lanczos vector. 
With the metric ansatz (50) we evaluate the following quantities 
in the most general forms
\begin{eqnarray}
{\cal R}_{GB}^2 &=&(p-1)p(p+1)\Big[(p+2)(U')^4 -4(U')^2 U''\Big]+2p(p+1)(n-1)
\big\{np+2(p+1)(n-2)\big\}\nn\\
&&(U')^2(V')^2+4p(p+1)(n-1)\Big[(p+1)(U')^3 V'- 2U'U''V'-p(U')^2V''\Big]
-4(n-1)\nn\\
&&(n-2)(p+1)\Big[U''(V')^2 + 2U'V'V''-(n-1)U'(V')^3 
+\big\{\frac{(p+2)}{2}(U')^2 -U''\big\}e^{2V}\Big]\nn\\
&&+2(n-1)(n-2)(n-3)\Big[2V''-(n-2)(V')^2-2(p+1)U'V'\Big] e^{2V}\nn\\
&&+(n-1)(n-2)(n-3)(n-4) e^{4V}
\end{eqnarray}
\begin{eqnarray}
Y_t^t&=&(p-1)p\Big[(p+1)(U')^4-3(U')^2U''\Big]+
(n-1)\big\{(n-2)(3p+1)+2p\big\}(U')^2(V')^2\nn\\
&&-4p(n-1)U'U''V'+\frac{p(n-1)}{2}\Big[2(3p+1)(U')^3V'-(p+1)(U')^2V''\Big]\nn\\
&&-(n-1)(n-2)\Big[U''(V')^2+(n-2)U'(V')^3+\frac{(p+1)}{2}U'V'V''\Big]\nn\\
&&+(n-1)(n-2)\Big[U''-(p+1)(U')^2-(n-3)U'V'\Big] e^{2V}=-Y_{x_i}^{x_i}
\end{eqnarray}
\begin{eqnarray}
Y_y^y&=&(p-1)p(p+1)\Big[(U')^4-(U')^2U''\Big]
+(n-1)(p+1)(n+p-2)(U')^2(V')^2\nn\\
&&+p(p+1)(n-1)\Big[2(U')^3V'-2U'U''V'-(U')^2V''\Big]+(n-1)(n-2)(p+1)\nn\\
&&\Big[U'(V')^3 -2U'V'V''-U''(V')^2+\big\{U''-(U')^2\big\}e^{2V}\Big]
+(n-1)(n-2)\nn\\
&&(n-3)\Big[(V')^2V''+(V')^4+\big\{V''+(V')^2\big\}e^{2V}\Big]
\end{eqnarray}
\begin{eqnarray}
Y_{\theta}^{\theta}&=&p(p+1)\Big[(p+1)(U')^3V'-(U')^2V''-2U'U''V'\Big]
+(p+1)(n-2)\nn\\
&&\Big[(3n-5)U'(V')^3-2U''(V')^2-4U'V'V''+\big\{2U''-(p+2)(U')^2\big\}e^{2V}
\Big]\nn\\
&&+(p+1)\big\{p(n-1)+2(n-2)(p+1)\big\}(U')^2(V')^2+(n-2)(n-3)\nn\\
&&\Big[\big\{3V''-(2n-5)(V')^2-3(p+1)U'V'\big\}e^{2V}-3V''(V')^2+
(n-1)(V')^4\Big]
\end{eqnarray}
\subsection{Planar Wall Solution}
   In this subsection, we solve the higher derivative field equations for 
$n=1$. These are the generalization of the well-known domain 
wall solution but includes contribution of higher-curvature terms 
into the field equations. Since the brane that we are considering is 
characterized by a conformally flat metric, we shall call the $p$-brane 
object a planar wall (i.e. $\hat{R}_{(4)}=0$) embedded in the bulk $AdS_5$. 
With the metric (50), the field equations (2) reduce to the form 
\begin{eqnarray}
\frac{p(p+1)}{2}u^2-\frac{(p-2)(p-1)p(p+1)}{2}\alpha\kappa^2 u^4
&=&-\kappa^2\big(\Lambda_b-\rho_0(y)\big)\\
\frac{p(p+1)}{2}u^2-\frac{(p-2)(p-1)p(p+1)}{2}\alpha\kappa^2 u^4
&=&-\kappa^2\big(\Lambda_b-\rho_y(y)\big)
\end{eqnarray}
Clearly, we see that the planar wall solution, i.e., $n=1$ and 
$\hat{R_{(4)}}=0$, requires the brane source terms in the extra 
dimensions to be equal, i.e., $\rho_o(y)=\rho_y(y)$ or they are trivial. 
In the former case we obtain 
\begin{equation}
 u^2=\frac{1\pm\sqrt{1+8\alpha\kappa^4(\Lambda_b-\rho_y)(p-1)(p-2)/{p(p+1)}}}
{2\alpha\kappa^2(p-1)(p-2)}, ~~ p > 2
\end{equation}
whose metric solution is given by (33). For the $3$-brane both the 
positive and negative root solutions are allowed 
if the inequality $0\leq\alpha\kappa^4(\rho_y-\Lambda_b)\leq3/4$ holds. 
This also does not rule out the positive bulk cosmological constant. 
Clearly, unlike the case in domain wall solution, where $\rho_0=\Lambda_b$ 
implies both $\hat{R}_{(4)}=0$ and $u=0$, the planar wall solution 
corresponds to $u=0 ~or ~M_{(5)}^3/{16\pi\alpha}$. 
We remind the reader that the case $\alpha=0$ does not correspond
to the Einstein gravity as long as the $3$-brane we are considering 
remains flat.
\subsection{Planar String Solution}
Now we turn to the case of co-dimensions two, i.e., $n=2$. With the 
ans$\ddot{a}$tze (23) and (50), the field equations (2) then reduce to the form
\begin{eqnarray}
-\Big(1-\frac{a(p+1)}{2(p-1)}u^2\Big)V''+(1-au^2)\big\{(V')^2+puV'\big\}
&+&\frac{p(p+1)}{2}u^2\nn\\
-\frac{(p-2)(p+1)}{4}a u^4&=&-\kappa^2(\Lambda_b-\rho_0)\\
(p+1)(1-au^2)uV'+\frac{p(p+1)}{2}u^2-\frac{(p-2)(p+1)}{4}a u^4
&=&-\kappa^2(\Lambda_b-\rho_y)\\
\frac{(p+1)(p+2)}{2}u^2-\frac{(p+1)(p+2)}{4}a u^4
&=&-\kappa^2(\Lambda_b-\rho_{\theta})
\end{eqnarray}
where $a=2p(p-1)\alpha\kappa^2$. We solve these equations for the physically 
interesting case of $3$-brane and obtain, whose general metric solution is 
given by (44) with $\hat{g}_{\mu\nu}=\eta_{\mu\nu}$, the following 
coefficients
\begin{eqnarray}
u^2&=&\frac{1\pm\sqrt{1+a\kappa^2(\Lambda_b-\rho_{\theta})/5}}{a}\nn\\
V(y)&=&\frac{uy}{b}+\frac{\alpha\kappa^4u}{10}
\int\frac{(4\Lambda_b-5\rho_y+\rho_{\theta})}
{b(1+b)}dy\nn\\
\end{eqnarray}
where $b=\sqrt{1+a\kappa^2(\Lambda_b-\rho_{\theta})/5}$ and 
$a=12\alpha\kappa^2$. These coefficients should also satisfy the following 
identity, derived from the equations (60-62), 
\begin{equation}
b\big\{(V')^2-V''-U'V'\big\}+\sqrt{5}\kappa^2(\rho_0-\rho_y)=0
\end{equation}
For simplicity one can choose the positive root of $u^2$ in (63). 
Here one can look for the solutions with a specific tuning condition as 
$\rho_y=\rho_{\theta}=\Lambda_b$. This means that the brane sources are 
trivial along with $\Lambda_b=0$ or they are the constant sources 
in the extra dimensions. In this case one gets $b=1$ and the general metric 
solution is given by (48) with $\eta_{\mu\nu}=\hat{g}_{\mu\nu}$. 
However, for the case of trivial brane sources in the extra dimensions, 
($\rho_i=0$), but with $\Lambda_b\neq 0$, one has an extra warp factor 
$c$, i.e.,
\begin{equation}
ds^2= e^{-2u|y|}\eta_{\mu\nu}dx^\mu dx^\nu + dy^2 
+ e^{-2u|y|c} d\theta^2
\end{equation}
where $c=\beta+1/b+2\alpha\kappa^4\Lambda_b/{5b(1+b)}$, $\beta$ is an 
integration constant, and 
$b=\sqrt{1+12\alpha\kappa^4\Lambda_b/5}$. Obviously, for the vanishing bulk 
cosmological constant (65) reduces to (48) with 
$\eta_{\mu\nu}=\hat{g}_{\mu\nu}$.
\nopagebreak[12]
\section{Dilatonic Gauss-Bonnet Gravity}
\subsection{Dilaton in the Brane Background}
In this section we consider the string inspired higher 
derivative gravity action in the dilatonic Gauss-Bonnet combination and 
study its compatibility with the RS type non-factorizable metrics (50) and 
the string amplitude computations, in particular the conformal weights of 
the dilaton couplings to the higher-curvature terms. 
The above mentioned action in the low energy limit and for the vanishing 
torsion, by rescaling dilaton to get the standard normalization of 
propagator correction-free Gauss-Bonnet scheme \cite{AAT}, may be written in 
$D(=d+1)$-dimensional space-time as
\begin{eqnarray}
S_M&=&\int d^{d+1}x \sqrt{-g}\big\{R-\epsilon(y)e^{m\Phi}+\lambda'e^{n\Phi}
\alpha(R^2-4R_{AB}R^{AB}+R_{ABCD}R^{ABCD})\nn\\
&&-\frac{4}{d-1}(\nabla\Phi)^2-V(\Phi)+\cdots\big\}
+\int_{\partial B} d^dx\sqrt{-g_{brane}}~ e^{q\Phi(y)}~ \tilde{\sigma}
\end{eqnarray}
where the last term is the contribution of boundary action defined
at the position of $p$-brane, $\lambda'=\alpha'/{8g_s^2}>0$, $\alpha'$ is 
the Regge slope, $g_s$ is the string coupling constant, $\tilde{\sigma}$ 
is the brane tension, $V(\Phi)$ is the dilatonic potential, which and the 
subsequent terms denoted by dots we set zero in the present 
consideration \footnote{ When this paper was being typed 
the reference \cite{NIC} with similar effective action appeared in the lanl 
e-Print archive where the form of potential 
$-V(\Phi)=c_2 f(\Phi)(\nabla_A\Phi)^4+\cdots$ is also included}. For 
generality one can keep the conformal parameters $m, n$ and $q$ arbitary in 
the higher derivative gravity coupled to dilaton. Nevertheless, 
$m$ and $n$ can be fixed for the particular space-time 
dimensions of the string frame by matching the coefficients with the 
tree-level amplitudes to ${\cal O}(\alpha')$ when one restricts the action 
(66) to the effective string amplitude computations. In the 
lower dimensions, $d<10$, the term $\epsilon(y)$ may be realized as the 
negative of the bulk cosmological constant. As explored in \cite{JMC} the 
RS type brane solutions can not be obtained in the low energy supergravities 
coming from string theory. A possibility left open in \cite{JMC} 
(see also the references therein) that the higher derivative corrections 
to the gravity action like the ones present in string theory or $M$-theory 
should allow the brane world solutions can be materialized if one defines 
$\epsilon(y)=-\Lambda_b$. In this sense the term $\epsilon(y)e^{m\Phi}$ is 
crucial in the obtention of RS type brane solutions and in making 
the conformal weight of dilaton couplings compatible with the string 
amplitude computations. In what follows we 
restrict the action (66) in the $d+1=5$-dimensional space-time.       
 
The graviton equation of motion derived from the action (66) can be 
expressed in the form, defining $e^{n\Phi(y)}=f(\Phi)$,
\begin{eqnarray}
0&=&G_{AB}+\half g_{AB}~\epsilon(\Phi)e^{m\Phi}
-\half\alpha\lambda' f(\Phi)\big(g_{AB}{\cal R}_{GB}^2-4Y_{AB}\big)
-\frac{4}{3}\big(\nabla_A\Phi\nabla_B\Phi-\half g_{AB}(\nabla\Phi)^2\big)\nn\\
&&+2\alpha\lambda'\Big\{R(g_{AB}\nabla^2-\nabla_A\nabla_B)f(\Phi)-2g_{AB}
\nabla^C\nabla^Df(\Phi) R_{CD}-4\nabla_C f(\Phi)\nabla^C R_{AB}\nn\\
&&-2\nabla^2f(\Phi)R_{AB} + 4\nabla_{(A} \nabla_{|C|} f(\Phi) R_{B)}\,^C
+4\nabla_Cf(\Phi)\nabla_{(A} R_{B)}\,^C
+2\nabla^{(C}\nabla^{D)}f(\Phi)R_{ACBD}\Big\}\nn\\
&&-e^{q\Phi(y)}\frac{\sqrt{-g_b}}
{2\sqrt{-g}}\delta_A^{\mu}\delta_B^{\nu}(g_b)_{\mu\nu}~ \sigma(y),
\end{eqnarray}
while the dilaton equation of motion is given by 
\begin{eqnarray}
0=\alpha\lambda' f'(\Phi){\cal R}_{GB}^2+\frac{8}{3}\nabla^2\Phi-m\epsilon(y) 
e^{m\Phi}+qe^{q\Phi}\frac{\sqrt{-g_b}}{\sqrt{-g}}\delta_A^{\mu}\delta_B^{\nu}
(g_b)_{\mu\nu} \sigma(y)
\end{eqnarray}
In eqn(68) prime denotes the differentiation w.r. to $\Phi$, ${\cal R}_{GB}^2$
 is the Gauss-Bonnet term, $Y_{AB}$ was defined previously in the section 5 
and $\sigma(y)=\tilde{\sigma} \delta(y)$. With the metric ansatz (50), 
for $d+1=5$, the field equations (67-68) reduce to the following forms
\begin{eqnarray}
12\lambda'\alpha e^{n\Phi}\big\{(U')^4-(U')^2U''-3n(U')^3U''
+n^2(U')^2(\Phi')^2+2nU'U''\Phi'+n(U')^2\Phi''\big\}\nn\\
~~~~~~~~~~~~-6(U')^2+3U''-\frac{2}{3}(\Phi')^2-\frac{1}{2} 
\epsilon(y)e^{m\Phi}+\frac{1}{2} e^{q\Phi}\sigma=0\\
12\lambda'\alpha e^{n\Phi}\big\{(U')^4-4n(U')^3\Phi'\big\}
-6(U')^2+\frac{2}{3}(\Phi')^2-\half\epsilon(y) e^{m\Phi}=0\\
24n\lambda'\alpha e^{n\Phi}\big\{5(U')^4-4(U')^2U''\big\}+\frac{8}{3}
\Phi''-\frac{32}{3}U'\Phi'-m\epsilon(y) e^{m\Phi}+qe^{q\Phi}\sigma(y)=0
\end{eqnarray}
Here the primes denote differentiation w.r. to the fifth bulk coordinate$(y)$.
 Notice that for the case of constant dilaton field w.r. to $y$ 
(i.e.,$\Phi(y)=\Phi_0, \Phi'(y)=\Phi''(y)=0$) and the bulk geometry having 
a warp factor $U(y)=u|y|$, where $u$ is some constant function, the 
first two equations in the bulk reduce to a single equation, i.e.
\begin{equation}
-12u^2+24\lambda'\alpha e^{n\Phi_0}u^4-\epsilon(y) e^{m\Phi_0}=0\nn
\end{equation}
This implies that in the bulk because of the Bianchi identities there 
could only be two independent equations. To see this explicitly from the 
equations (69-71) one can formally arrive at the following identity (see 
also \cite{NIC})  
\begin{equation}
e^{q\Phi(y)}\sigma(y)\big\{4U'(y)-q\Phi'(y)\big\}-e^{m\Phi}\epsilon'(y)=0
\end{equation}
When one defines $\epsilon(y)(=-\Lambda_b)$ as the bulk cosmological 
constant, then the requirement that $\epsilon'(y)=0$ equally avoids the 
breaking of Poincare invariance in the bulk \cite{RS1,VAR,NIC}.
\subsection{Constant Dilaton Solution}
One can solve the above field equations with the constant 
dilaton in the $AdS$ bulk i.e. the dilaton remains constant w.r. to the 
radial bulk coordinate, $\Phi(y)=\Phi_0$. This can also be associated with 
the $D3$ brane solution of certain version of the string theories. For this 
case, the Bianchi identity (72) is reduced to 
\begin{equation}
4e^{q\Phi_0}U'(y)\tilde{\sigma}\delta(y)=0\nn
\end{equation}
which is trivial since $\delta(y)=0$ in the bulk. And the field equations 
(69-70) reduce to the following forms
\begin{eqnarray}
\frac{d}{dy}\big(8\lambda'\alpha (U')^3 e^{n\Phi_0}-6U'\big)
&=&e^{q\Phi_0}\tilde{\sigma} \delta(y)\\
\epsilon(y)e^{m\Phi_0}+12(U')^2-24\lambda'\alpha (U')^4 e^{n\Phi_0}&=&0
\end{eqnarray}
where in obtaining eqn(73) we have utilized the equation (74). In the bulk 
eqn(73) implies that
\begin{equation}
8\lambda'\alpha (U')^3 e^{n\Phi_0}-6U'= c
\end{equation}
where $c$ is an integration constant. For the constant dilaton the 
additional terms $-V(\Phi)=c_2f(\Phi)(\nabla_A(\Phi))^4\cdots$ considered in 
\cite{NIC} does not affect this equation. As a third-degree equation, (75) 
must have atleast one real solution for the arbitary constant $c$. 
Indeed, it gives all three real solutions if we impose $a\leq c/2\leq -a$, 
where $a=e^{-n\Phi_0/2}/\sqrt{\lambda'\alpha}$. However, to make the various 
conformal parameters: m, n, q compatible with the string 
amplitude computations, one needs to restrictively define the integration 
constant as  $c=8 e^{-n\Phi_0/2}/{3\sqrt{3\lambda'\alpha}}$ with some 
suitable junction conditions about the brane position. With this one of the 
real solutions of eqn(75) is given by 
\begin{equation}
U'(y)=k_+=-k_-=k=\frac{e^{-n\Phi_0/2}}{2\sqrt{3\lambda'\alpha}}
\end{equation}
Obviously, if $y>0$ corresponds to the solution $U'(y)=k_+$, then $y<0$ 
corresponds to $U'(y)~=-k_-$. The above choice of $c$ uniquely reproduces 
the conformal weight of dilaton 
couplings compatible with string amplitude computations. Indeed, 
(76) can also be associated to the RS solution. For the vanishing dilaton 
field this implies $k=k_+=-k_-=1/{(2\sqrt{3\lambda'\alpha})}$. 
Before going to make any conclusion here, we first 
give the full treatment of all equations both in the bulk and on 
the brane and find the general and consistent solutions.

Again, integrating the equation (73) over 
an interval that includes the brane position at $y=0$, one gets
\begin{equation}
e^{q\Phi_0}\tilde{\sigma}
=-6(k_+-k_-)+8\lambda'\alpha (k_+^3-k_-^3) e^{n\Phi_0}
\end{equation}
This relates the $k_+,~ k_-$ with the brane tension $\tilde{\sigma}$. 
Now, one needs to solve eqn(74) with the continuity of 
bulk cosmological constant $\epsilon(y)(=-\Lambda_b(y))$ at the brane 
position $y=0$. Therefore, from (74) one has
\begin{equation}
\epsilon(y)e^{m\Phi_0}=12k_{\pm}^2(2\lambda'\alpha k_{\pm}^2 e^{n\Phi_0}-1)
\end{equation}
implying
\begin{equation}
\epsilon(y) e^{m\Phi_0}=12 k_+^2(2\lambda'\alpha k_+^2 e^{n\Phi_0}-1)
=12 k_-^2(2\lambda'\alpha k_-^2 e^{n\Phi_0}-1)
\end{equation}
This equation is solved for $(i)~ k_+=\pm k_-=k$ and  
(ii) $k_+^2+k_-^2=e^{-n\Phi_0}/{2\lambda'\alpha}$. Indeed, the second 
solution covers a broad region of the solution space, 
where the first solution could be a point on that space. 
Ofcourse, the solution with $k_+=k_-$ can not be the RS solution as this 
has continuous metric function at $y=0$ which violates the symmetry about 
the brane position. So we choose the negative sign in the first solution. 
Notice that for the constant $c=2 e^{-n\Phi_0/2}/\sqrt{\lambda'\alpha}$, 
eqn(75) can have two real solutions as $k_{\pm}=\pm e^{-n\Phi_0/2}/
{\sqrt{\lambda'\alpha}}$ and $k_{\pm}=\pm e^{-n\Phi_0/2}/
{2\sqrt{\lambda'\alpha}}$ which satisfy both the solutions of (79) 
independently. If this is the case, the theory is compatible only 
with RS type space-time but not with the string amplitudes computation. 
 The solution which is compatible with both the RS type geometry and the 
string theory is given by (76). And, a  possible pair of imaginary solutions 
is discarded since this demands $k_+^2+k_-^2<0$ and hence $\lambda'<0$ for 
positive $\alpha$, but this is again not compatible with the string amplitude 
computations. Therefore, for the choice $k_+=-k_-=k$, one gets
\begin{equation}
\epsilon(y)=12k^2(2\lambda'\alpha k^2 e^{n\Phi_o}-1) e^{-m\Phi_0}
\end{equation}
   Now we solve the equation (71) in the bulk, where one can set 
$\Phi'(y)=\Phi''(y)=0$, but proper care should be taken at the brane 
position, i.e. $y=0$, where discontinuity of $\Phi(y)$ may appear due to 
the delta function source. Eqn(71), in conjuction with eqn(74) and 
$U''(y)=0$, reduces to the form
\begin{equation}
(5n-m)\epsilon(y)e^{m\Phi_0}+60n k_{\pm}^2
+q e^{q\Phi_0}\tilde{\sigma}(y)\delta(y)=0\nn
\end{equation}
which, in the bulk, on using (78) reduces to the form
\begin{equation}
2(5n-m)\lambda'\alpha k_{\pm}^2 e^{n\Phi_0}+m=0
\end{equation}
For the solution $k_+=-k_-=k$ one has
\begin{equation}
m=\frac{10n\lambda'\alpha k^2 e^{n\Phi_0}}{2\lambda'\alpha k^2 e^{n\Phi_0}-1}
\end{equation}
If we integrate eqn(71) in the neighbourhood of brane 
position, $y=0$, with $U'(y)=k_+=-k_-$, one is left with
\begin{eqnarray}
q \int_{0_-}^{0_+}e^{q\Phi_0}\tilde{\sigma}\delta(y)
&=&\int_{0_-}^{0_+}\frac{d}{dy}\big(32n\lambda'\alpha e^{q\Phi_0}(U'(y))^3\big)dy\nn\\
i.e.~~q e^{q\Phi_0}\tilde{\sigma}&=&32n\lambda'\alpha e^{n\Phi_0}
(k_+^3-k_-^3) 
\end{eqnarray} 
Eqn(83) in conjuction with eqn(77) implies
\begin{equation}
\tilde{\sigma}=\frac{48nk}{q-4n} e^{-q\Phi_0}
\end{equation}
where 
\begin{equation}
q=\frac{16nk^2\lambda'\alpha}{4k^2\lambda'\alpha e^{n\Phi_0}-3}
\end{equation}
Notice that in the string amplitude computations \cite{ESF,AAT}, the 
conformal parameters $m$ and $n$ are fixed for the given space-time 
dimensions of the string frame. 
In $d+1=5$ dimensions, $m= 4/{(d-1)} = 4/3 = -n $. With these 
conformal weights and vanishing dilaton field, from equations (82), (85), 
(84) and (80) we get (see also \cite{NIC})
\begin{equation}
k=\frac{1}{2\sqrt{3\lambda'\alpha}}, ~ q=\frac{2}{3},~ 
\tilde{\sigma}(y)=-\frac{32 k}{3}, ~ \Lambda_b=-\epsilon(y)=10 k^2
\end{equation}
where $\lambda'=1/{8 g_s^2}$. These values are well consistent with both the 
RS brane world scenario and 5-dimensional string theory amplitudes. In the 
former case the bulk cosmological constant $\Lambda_b (=10k^2)$ appears to 
have opposite sign than that of the brane tension $\tilde{\sigma} (=-32k/3)$ 
to ensure the positive tension brane at $y=0$, and in the latter case the 
required values and relations among the conformal weights of dilaton coupling 
in the five dimensional string theory are satisfied with the proper choice of 
$\lambda'$ or $\alpha$ or both. 

A possible solution of eqn(75) as 
\begin{equation}
k_+=-k_-=k=\frac{e^{-n\Phi_0/2}}{2\sqrt{\lambda'\alpha}}, 
\end{equation}
can be associated to the RS solution. This also reproduces 
the correct sign of conformal weight of the dilaton couplings as in the 
string amplitude computations. One has in this case 
\begin{equation}
m=-5n=\frac{4}{3}, ~~  q=\frac{8}{15}~~ \tilde{\sigma}
=-8k,~~\Lambda_b=6k^2 
\end{equation}           
This may appear to contradict the second solution of (79) in the sense 
that $k_+^2+k_-^2= e^{-n\Phi_0}/{2\lambda'\alpha}$ for $k_+=-k_-$. But, 
we stress that this need not be the case. As we mentioned 
above that the solution-space of the second solution of (79) is broad and it 
can cover the first solution as a particular point in that space, where the 
$p$-brane could be located. Further, with the choice of a solution as  
$k_+=-k_-= e^{-n\Phi_0/2}/{\sqrt{\lambda'\alpha}}$, one gets  
$m= 10,  n= 4/3,  q= 32/15,  \tilde{\sigma} = 4k,  \Lambda_b= -12k^2$. 
Here the sign of the brane tension and the bulk cosmological constant get 
reversed, thereby implies some possible de Sitter region in the solution 
space. However, since the sign of the conformal weight $m$ is same that of 
$n$, these values are not acceptable from the view point of string 
amplitude computations.        

Finally, for the solution 
$k_+^2+k_-^2=e^{-n\Phi_0}/{2\sqrt{\lambda'\alpha}}$, but 
$k_+\neq -k_-$, one has from eqn(81) that $m+5n=0$ and 
$m(1-4\lambda'\alpha e^{n\Phi_0} k_{\pm}^2)=0$. This implies that $m=-5n=0$, which from 
eqn(83) further implies that $q=0$. For the trivial dilaton couplings 
these results somewhat contradict to those of ref.\cite{KKL}, where there 
exists some dependence of $k_{\pm}$ on $\epsilon(y)(=-\Lambda_b)$ for 
non-trivial $\alpha$ and $\lambda'$. Similar conclusions have recently 
been made in \cite{NIC} and the solutions for linear dilaton with the 
trivial case $m=n=q=0$ is also generalized. However, there are no solutions 
of the bulk equation (75) other than mentioned above but still holds 
the equality $k_+^2+k_-^2=e^{-n\Phi_0}/{2\sqrt{\lambda'\alpha}}$. 
Hence, we are in a dilema if the two solutions of (79) are linearly 
independent. The case of linear dilaton with $m=n=q=0,~ V(\Phi)=0$ and 
$\epsilon(y)={\cal V}(\Phi)$, dilaton super potential is studied in 
\cite{ILZ}.
\subsection{Linear Dilaton: Time Dependent Solution}
Our starting point for this subsection is also the string-inspired low 
energy effective action in the higher derivative gravity model without 
torsion. However, unlike in the previous subsection, here we assume a simple 
dilaton-gravity coupling also with the curvature scalar ($R$). Indeed, 
this is the case one has after 
rescaling $g_{AB}$ in order to pass to the scheme in which the whole 
effective lagrangian is multiplied by $e^{-2\Phi}$ and the term 
$c_2(\nabla\Phi)^4$ is formally absent in the low energy effective action, 
a scheme often known as `$\sigma$-parametrization'. This is also 
necessary for the correspondence with Weyl invariance of the action. 
Confining ourselves to the $5$-dimensional spacetime,  
\begin{equation}
S_M=\int d^5x\sqrt{-g}e^{-2\Phi}\Big\{
R+\lambda'\alpha\big(R^2-4 R_{AB}R^{AB}+R_{ABCD}R^{ABCD}\big)
-4(\nabla\Phi)^2-V(\Phi)\Big\}
\end{equation}
where $V(\Phi)$ is the dilatonic potential which we set zero. The 
inflationary solutions to the above action by considering a conformally flat 
Friedmann-Robertson-Walker metric were realized in \cite{ILS}, where the 
extra dimensions were equally treated as large as the usual 3 spatial 
dimensions. But this is not the case we are considering. We assume the 
background metric in the following form
\begin{equation}
ds^2 = - d\tau^2+e^{-2U(\tau, y)}\delta_{ij}dx^idx^j+e^{-2W(\tau, y)} 
(dr^2+r^2d\Omega_{n-1}^2)
\end{equation}
where $d\tau$ is the conformal time interval, so that the usual $3+1$ 
dimensional metric can be expressed in the form 
$e^{-2U(t, y)}\eta_{\mu\nu}dx^{\mu}dx^{\nu}$, the indices $i,j$ denote 
the spatial indices $1,2,3$. We solve the field equations in 
$D=5$-dimensional space-time with a specific choice of the ansatz 
$U^{-1}(\tau, y)=W(\tau, y)$. This choice may be realized for the 
inflationary solutions with the expanding external space and contracting 
internal spatial dimensions, which should also be true in brane set up 
where as the brane inflates the extra bulk coordinates get contracted. 
However, this may not be the case with stable non-compact extra dimensions. 
We concentrate only on the case with $n=1$. Assuming the usual 3-dimensional 
space as homogeneous and isotropic, the 5-dimensional effective action 
reduces to the form:
\begin{equation}
S_M=\int d\tau e^{-2\Phi} e^{-2U}\Big\{2H^2+\dot{H}+\dot{\Phi}^2
-6\lambda'\alpha\big(2\dot{H}H^2+H^4\big)\Big\}
\end{equation}
where dot denotes the differentiation $\partial/{\partial \tau}$ and we 
have defined $-\dot{U}(\tau)= H(\tau)$, the Hubble parameter. The 
corresponding field equations for the scale factor and dilaton field are
\begin{equation}
\ddot{\Phi}-\dot{\Phi}^2+ 2H^2+\dot{H}+2H\dot{\Phi}
-6\lambda' \alpha (2\dot{H}H^2+H^4)=0
\end{equation}
\begin{equation}
\ddot{\Phi}-3\dot{\Phi}^2+6\lambda'\alpha\big\{H^4+2H^2\dot{H}
-4\dot{\Phi}H^3-4H\dot{H}\dot{\Phi}-2H^2(\ddot{\Phi}-2\dot{\Phi}^2)\big\}=0
\end{equation}
Indeed, one can modify the action (89) by adding a boundary term so that
it does not contain the second derivative of $H$. However, here we solve 
the above two equations for a linear dilaton field with the constant
Hubble parameter, i.e. $\dot{\Phi}=\upsilon= constant$, \,$H=H_0= constant$. 
and obtain the following exact real solution:\\
(i) the trivial solution corresponds to $H_0=0,\,\dot{\Phi}=0$, and 
(ii) the non-trivial real solution implies 
\begin{equation}
H_0^2=\frac{1}{8\lambda'\alpha}, ~~ \frac{1}{2\lambda'\alpha}
\end{equation}
\begin{equation}
\upsilon=\big\{1\pm\sqrt{3(1-2\lambda'\alpha H_0^2)}\big\}H_0
\end{equation}
There could be a pair of imaginary solutions both to $H_0$ and $\upsilon$, 
but we discard them. The above solutions could be physically more 
meaningful as compared to the solutions found in \cite{ILS}, in the sense 
that one needs either of the coupling constants $\lambda'$ or $\gamma_3$ 
(see the ref.\cite{ILS}) to be negative. If one judges the action from 
string low energy effective action and the required relations among 
Gauss-Bonnet couplings $\alpha, \beta, \gamma$, a somewhat unphysical 
relation persists, otherwise in their solution the Hubble 
parameter becomes imaginary but this is not the case in our solution. 
Further, the metric ansatz chosen in \cite{ILS} is conformally flat FRW 
type and is somewhat unphysical to explain the inflationary solutions, in the 
sense that the extra spatial dimensions equally inflate during the 
inflationary epoch. But in our parametrization, the extra spatial dimensions 
get contracted when the usual 3-brane world volume expands.   

\section{Discussions and Outlook}
The background metric with the Ricci constant curvature would imply  a 
constraint equation in the theory of higher derivative gravity which defines  
the bulk space-time geometry as $AdS(dS)$. We have studied various solutions 
of the higher derivative gravity in the brane background by considering the 
string inspired effective actions in the Gauss-Bonnet combination. The defect 
solutions in higher dimensions known to the Einstein equations are 
generalized by including the contribution of higher-curvature terms into 
the field equations and the general metric solutions that correspond to 
planar defects are obtained. The string inspired higher derivative gravity 
theory in the dilatonic Gauss-Bonnet combination is shown to be consistent 
to both the RS type non-factorizable geometry and the conformal weights of 
dilatonic couplings in the string amplitude computations with the proper 
choice of free parameters of the theory. Discussed is also time dependent 
dilaton solutions in a version of string inspired higher derivative gravity 
model coupled to dilaton.
 
There are few interesting topics which deserve future investigations. One 
of these is to extend the sections 5 and 6 to explain 
the different brane-world black-hole (BH) solutions without and with dilaton. 
One such solution as Schwarzschild $AdS_5$ black-hole in the background of 
constant Ricci curvature space-time is realized in ref. \cite{SNS}. 
It is also important to further explore dilatonic Gauss-Bonnet BH solutions 
with more general metric ansatz in the curved and flat brane background. It 
is also interesting to investigate the cosmological implications of the 
higher derivative gravity theory of the type considered in Section 6.1-6.2. 
\section*{Acknowledgements}
The author is greatful to Prof. Y. M. Cho for the constant 
encouragement and he would also like to thank S.-H. Moon for the 
discussions. The author is greatful to I. Antoniadis, C. Nu$\tilde{n}$ez, 
and S. Odintsov for the useful comments and important remarks in the 
previous version of the manuscript. This work is supported in part by the 
BK21 project of the Ministry of Education, Korea.

\end{document}